\documentclass[aps,preprint]{revtex4}%
\usepackage{amsfonts}
\usepackage{amsmath}
\usepackage{amssymb}
\usepackage{graphicx}%
\begin{document}
\title{Higher dimensional Yang-Mills black holes in third order Lovelock gravity}
\author{S. Habib Mazharimousavi$^{\ast}$}
\author{M. Halilsoy$^{\dag}$}
\affiliation{Department of Physics, Eastern Mediterranean University,}
\affiliation{G. Magusa, north Cyprus, Mersin-10, Turkey}
\affiliation{$^{\ast}$habib.mazhari@emu.edu.tr}
\affiliation{$^{\dagger}$mustafa.halilsoy@emu.edu.tr}

\begin{abstract}
By employing the higher (N%
$>$%
5) dimensional version of the Wu-Yang Ansatz we obtain magnetically charged
new black hole solutions in the Einstein-Yang-Mills-Lovelock (EYML) theory
with second ($\alpha_{2}$) and third ($\alpha_{3}$)order parameters. These
parameters, where $\alpha_{2}$ is also known as the Gauss-Bonnet parameter,
modify the horizons (and the resulting thermodynamical properties) of the
black holes. It is shown also that asymptotically ($r\rightarrow\infty$),
these parameters contribute to an effective cosmological constant -without
cosmological constant- so that the solution behaves de-Sitter (Anti de-Sitter) like.

\end{abstract}
\maketitle

\section{\bigskip INTRODUCTION}

As the requirements of string theory/brane world cosmology, higher dimensional
($N>4$) space times have been extensively investigated during the recent
decades. Extensions of $N=4$ Einstein's gravity has already gained enough
momentum from different perspectives. General relativity admits local black
hole solutions as well as global cosmological solutions such as de Sitter (dS)
and anti de-Sitter (AdS), which are important from the field theory (i.e.,
AdS/CFT) correspondence point of view. Once a pure gravity solution (with or
without a cosmological constant) is found the next (routine) step has been to
search for the corresponding Einstein-Maxwell (EM) solution with the inclusion
of electromagnetic fields. As a result EM dS (AdS) space times have all been
obtained and investigated to great extend. One more extension from a different
viewpoint, which is fashionable nowadays, is to consider extra terms in the
Einstein-Hillbert action such as the ones considered by Lovelock in higher
dimensions to obtain solutions, both black holes and cosmological \cite{1}.
The added extra terms in the action have the advantage, as they should, that
they don't give rise to higher order field equations. It is known that in
$N=4$, the second order (also known as Gauss-Bonnet) Lovelock Lagrangian
becomes trivial unless coupled with non-trivial sources such as non-minimal
scalar fields \cite{9}. In higher dimensions ($N\geq5$), however coupling with
electromagnetic fields proved fruitful and gave rise to interesting black
hole/cosmological solutions\cite{2}. In this regard, Brihaye, et al, have
worked on particle-like solutions of EYM fields coupled with GB gravity in
N-dimensional spherically symmetric spacetime \cite{5}. To have a non-trivial
theory with the higher order Lovelock Lagrangian with third order parameter on
the other hand, we need the dimensionality of our space time to be $N\geq7.$
In this paper we shall follow similar steps, to extend the results of
electromagnetic fields to the Yang-Mills (YM) fields with gauge group
$SO(N-1).$ As expected, going from Maxwell to YM constitutes a highly
non-trivial process originating from the inherent non-linearity of the latter.
To cope with this difficulty we employ a particular YM Ansatz solution which
was familiar for a long time to the high energy physics community. This is the
Wu-Yang Ansatz, which was originally introduced in $N=4$ field theory
\cite{3,4}. Recently we have generalized this Ansatz to $N=5$, in the
Einstein-Gauss-Bonnet (EGB) theory and obtained a new
Einstein-Yang-Mills-Gauss-Bonnet (EYMGB) black hole \cite{5}. By a similar
line of thought we wish to extend those results further to $N>5$ and also
within the context (for $N\geq7$) of third order Lovelock gravity. Our results
show that both the second ($\alpha_{2}$) and third ($\alpha_{3}$) order
parameters modify the EYM black holes as well as their formation
significantly. For instance in the $g_{tt}$ term the gauge charge term comes
with the opposite sign and fixed power of $\frac{1}{r^{2}}$ , which is
unprecedented in the realm of EM black holes of higher dimensions. This makes
construction of black hole types from pure YM charge (with negligible mass,
for example) possible and enriches our list of black holes with new
properties. What follows for $N\geq7,$ for technical reasons, we assume a
relation between $\alpha_{2}$ and $\alpha_{3}$ which are completely free
otherwise. Not only black holes but the asymptotical behaviors and properties
of our space times are determined by these parameters as well. It is not
difficult to \ anticipate that by the same token the same problem can further
be generalized to cover the forth ($\alpha_{4}$), fifth ($\alpha_{5}$), etc.
order terms in the action to be superimposed to the Einstein-Hillbert (EH)
Lagrangian. By studying the relative weight of contribution from higher
Lovelock terms it is not difficult to anticipate that the EH and GB terms
dominate over the higher order, much more tedious terms. For this reason we
restrict ourselves in this paper to maximum third order ($\alpha_{3}$) terms
in the Lagrangian.

\section{ACTION\ AND\ FIELD\ EQUATIONS}

The action which describes the third order Lovelock gravity coupled with
Yang-Mills field without a cosmological constant in $N$ dimension reads
\cite{1}%
\begin{equation}
I_{G}=\frac{1}{2}\int_{\mathcal{M}}dx^{N}\sqrt{-g}\left[  \mathcal{L}%
_{EH}+\alpha_{2}\mathcal{L}_{GB}+\alpha_{3}\mathcal{L}_{\left(  3\right)
}-tr\left(  F_{\mu\nu}^{(a)}F^{(a)\mu\nu}\right)  \right]  ,
\end{equation}
where $tr(\mathbf{.})=\overset{\left(  N-1\right)  (N-2)/2}{\underset
{a=1}{\sum}}\left(  \mathbf{.}\right)  ,$ $\mathcal{L}_{EH}=R$ is the
Einstein-Hilbert Lagrangian, $\mathcal{L}_{GB}=R_{\mu\nu\gamma\delta}R^{\mu
\nu\gamma\delta}-4R_{\mu\nu}R^{\mu\nu}+R^{2}$ is the Gauss-Bonnet (GB)
Lagrangian, and
\begin{align}
\mathcal{L}_{\left(  3\right)  }  &  =2R^{\mu\nu\sigma\kappa}R_{\sigma
\kappa\rho\tau}R_{\quad\mu\nu}^{\rho\tau}+8R_{\quad\sigma\rho}^{\mu\nu
}R_{\quad\nu\tau}^{\sigma\kappa}R_{\quad\mu\kappa}^{\rho\tau}\nonumber\\
&  +24R^{\mu\nu\sigma\kappa}R_{\sigma\kappa\nu\rho}R_{\ \mu}^{\rho}%
+3RR^{\mu\nu\sigma\kappa}R_{\sigma\kappa\mu\nu}\nonumber\\
&  +24R^{\mu\nu\sigma\kappa}R_{\sigma\mu}R_{\kappa\nu}+16R^{\mu\nu}%
R_{\nu\sigma}R_{\ \mu}^{\sigma}\\
&  -12RR^{\mu\nu}R_{\mu\nu}+R^{3},\nonumber
\end{align}
is the third order Lovelock Lagrangian. Here $R,$ $R_{\mu\nu\gamma\delta\text{
}}$ and $R_{\mu\nu}$ are the Ricci Scalar, Riemann and Ricci tensors
respectively, while the gauge fields $F_{\mu\nu}^{\left(  a\right)  }$ are
\begin{equation}
F_{\mu\nu}^{\left(  a\right)  }=\partial_{\mu}A_{\nu}^{\left(  a\right)
}-\partial_{\nu}A_{\mu}^{\left(  a\right)  }+\frac{1}{2\sigma}C_{\left(
b\right)  \left(  c\right)  }^{\left(  a\right)  }A_{\mu}^{\left(  b\right)
}A_{\nu}^{\left(  c\right)  },
\end{equation}
where $C_{\left(  b\right)  \left(  c\right)  }^{\left(  a\right)  }$ are the
structure constants of $\frac{\left(  N-1\right)  (N-2)}{2}$-parameter Lie
group $G,$ $\sigma$ is a coupling constant, $A_{\mu}^{\left(  a\right)  }$ are
the gauge potentials, and $\alpha_{2}$ and $\alpha_{3}$ are GB and third order
Lovelock coefficients. Variation of the action with respect to the space-time
metric $g_{\mu\nu}$ yields the Einstein-Yang-Mills-Gauss-Bonnet-Lovelock
(EYMGBL) equations%
\begin{equation}
G_{\mu\nu}^{E}+\alpha_{2}G_{\mu\nu}^{GB}+\alpha_{3}G_{\mu\nu}^{\left(
3\right)  }=T_{\mu\nu},
\end{equation}
where the stress-energy tensor is
\begin{equation}
T_{\mu\nu}=tr\left[  2F_{\mu}^{\left(  a\right)  \lambda}F_{\nu\lambda
}^{\left(  a\right)  }-\frac{1}{2}F_{\lambda\sigma}^{\left(  a\right)
}F^{\left(  a\right)  \lambda\sigma}g_{\mu\nu}\right]  ,
\end{equation}
$G_{\mu\nu}^{E}$ is the Einstein tensor, while $G_{\mu\nu}^{GB}$ and
$G_{\mu\nu}^{\left(  3\right)  }$ are given explicitly as\cite{2}%
\begin{equation}
G_{\mu\nu}^{GB}=2\left(  -R_{\mu\sigma\kappa\tau}R_{\quad\nu}^{\kappa
\tau\sigma}-2R_{\mu\rho\nu\sigma}R^{\rho\sigma}-2R_{\mu\sigma}R_{\ \nu
}^{\sigma}+RR_{\mu\nu}\right)  -\frac{1}{2}\mathcal{L}_{GB}g_{\mu\nu},
\end{equation}%
\begin{gather}
G_{\mu\nu}^{\left(  3\right)  }=-3\left(  4R_{\qquad}^{\tau\rho\sigma\kappa
}R_{\sigma\kappa\lambda\rho}R_{~\nu\tau\mu}^{\lambda}-8R_{\quad\lambda\sigma
}^{\tau\rho}R_{\quad\tau\mu}^{\sigma\kappa}R_{~\nu\rho\kappa}^{\lambda}\right.
\\
+2R_{\nu}^{\ \tau\sigma\kappa}R_{\sigma\kappa\lambda\rho}R_{\quad\tau\mu
}^{\lambda\rho}-R_{\qquad}^{\tau\rho\sigma\kappa}R_{\sigma\kappa\tau\rho
}R_{\nu\mu}+8R_{\ \nu\sigma\rho}^{\tau}R_{\quad\tau\mu}^{\sigma\kappa
}R_{\ \kappa}^{\rho}\nonumber\\
+8R_{\ \nu\tau\kappa}^{\sigma}R_{\quad\sigma\mu}^{\tau\rho}R_{\ \rho}^{\kappa
}+4R_{\nu}^{\ \tau\sigma\kappa}R_{\sigma\kappa\mu\rho}R_{\ \tau}^{\rho
}-4R_{\nu}^{\ \tau\sigma\kappa}R_{\sigma\kappa\tau\rho}R_{\ \mu}^{\rho
}\nonumber\\
+4R_{\qquad}^{\tau\rho\sigma\kappa}R_{\sigma\kappa\tau\mu}R_{\nu\rho}%
+2RR_{\nu}^{\ \kappa\tau\rho}R_{\tau\rho\kappa\mu}+8R_{\ \nu\mu\rho}^{\tau
}R_{\ \sigma}^{\rho}R_{\ \tau}^{\rho}\nonumber\\
-8R_{\ \nu\tau\rho}^{\sigma}R_{\ \sigma}^{\tau}R_{\ \mu}^{\rho}-8R_{\quad
\sigma\mu}^{\tau\rho}R_{\ \tau}^{\sigma}R_{\nu\rho}-4RR_{\ \nu\mu\rho}^{\tau
}R_{\ \tau}^{\rho}\nonumber\\
+4R_{\quad}^{\tau\rho}R_{\rho\tau}R_{\nu\mu}-8R_{\ \nu}^{\tau}R_{\tau\rho
}R_{\ \mu}^{\rho}+4RR_{\nu\rho}R_{\ \mu}^{\rho}\nonumber\\
\left.  -R^{2}R_{\nu\mu}\right)  -\frac{1}{2}\mathcal{L}_{\left(  3\right)
}g_{\mu\nu}.\nonumber
\end{gather}

Variation of the action with respect to the gauge potentials $A_{\mu}^{\left(
a\right)  }$ yields the Yang-Mills equations
\begin{equation}
F_{;\mu}^{\left(  a\right)  \mu\nu}+\frac{1}{\sigma}C_{\left(  b\right)
\left(  c\right)  }^{\left(  a\right)  }A_{\mu}^{\left(  b\right)  }F^{\left(
c\right)  \mu\nu}=0,
\end{equation}
while the integrability conditions are%
\begin{equation}
\ast F_{;\mu}^{\left(  a\right)  \mu\nu}+\frac{1}{\sigma}C_{\left(  b\right)
\left(  c\right)  }^{\left(  a\right)  }A_{\mu}^{\left(  b\right)  }\ast
F^{\left(  c\right)  \mu\nu}=0,
\end{equation}
in which * means duality\cite{6}.

\section{WU-YANG\ ANSATZ\ IN\ $N>5$\ DIMENSIONS}

The $N-$dimensional line element is chosen as
\begin{equation}
ds^{2}=-f(r)\;dt^{2}+\frac{dr^{2}}{f(r)}+r^{2}d\;\Omega_{N-2}^{2},
\end{equation}
in which the $S^{N-2}$ line element will be expressed in the standard
spherical form%
\begin{equation}
d\Omega_{N-2}^{2}=d\theta_{1}^{2}+\underset{i=2}{\overset{N-3}{%
{\textstyle\sum}
}}\underset{j=1}{\overset{i-1}{%
{\textstyle\prod}
}}\sin^{2}\theta_{j}\;d\theta_{i}^{2},\text{ }0\leq\theta_{1}\leq\pi
,0\leq\theta_{i}\leq2\pi.
\end{equation}

We use the Wu-Yang Ansatz \cite{7} in $N-$dimensional case as%
\begin{align}
A^{(a)}  &  =\frac{Q}{r^{2}}\left(  x_{i}dx_{j}-x_{j}dx_{i}\right)  ,\text{
\ \ }Q=\text{charge, \ }r^{2}=\overset{N-1}{\underset{i=1}{\sum}}x_{i}^{2},\\
2  &  \leq j+1\leq i\leq N-1,\text{ \ and \ }1\leq a\leq\left(  N-1\right)
(N-2)/2,\nonumber
\end{align}
where we imply (to have a systematic process) that the super indices $a$ is
chosen according to the values of $i$ and $j$ in order.

The YM field 2-forms are defined as follow%
\begin{equation}
F^{\left(  a\right)  }=dA^{\left(  a\right)  }+\frac{1}{2Q}C_{\left(
b\right)  \left(  c\right)  }^{\left(  a\right)  }A^{\left(  b\right)  }\wedge
A^{\left(  c\right)  }.
\end{equation}

We note that our notation follows the standard exterior differential forms,
namely $d$ stands for the exterior derivative while $\wedge$ stands for the
wedge product \cite{7}. The integrability conditions
\begin{equation}
dF^{\left(  a\right)  }+\frac{1}{Q}C_{\left(  b\right)  \left(  c\right)
}^{\left(  a\right)  }A^{\left(  b\right)  }\wedge F^{\left(  c\right)  }=0,
\end{equation}
are easily satisfied by using (12). The YM equations
\begin{equation}
d\ast F^{\left(  a\right)  }+\frac{1}{Q}C_{\left(  b\right)  \left(  c\right)
}^{\left(  a\right)  }A^{\left(  b\right)  }\wedge\ast F^{\left(  c\right)
}=0,
\end{equation}
are also satisfied. The energy-momentum tensor (5), becomes after%
\begin{equation}
tr\left[  F_{\lambda\sigma}^{\left(  a\right)  }F^{\left(  a\right)
\lambda\sigma}\right]  =\frac{\left(  N-3\right)  \left(  N-2\right)  Q^{2}%
}{r^{4}},
\end{equation}
with the non-zero components%
\begin{equation}
T_{\text{ }b}^{a}=\frac{\left(  N-3\right)  \left(  N-2\right)  Q^{2}}{2r^{4}%
}\text{diag}\left[  -1,-1,\kappa,\kappa,..,\kappa\right]  ,\text{ \ and
\ }\kappa=-\frac{N-6}{N-2}.
\end{equation}

The EYMGBL equations (4) reduce to the general equation%
\begin{gather}
\left(  r^{5}-2\tilde{\alpha}_{2}r^{3}\left(  f\left(  r\right)  -1\right)
+3\tilde{\alpha}_{3}r\left(  f\left(  r\right)  -1\right)  ^{2}\right)
f^{\prime}\left(  r\right)  +\nonumber\\
\left(  n-1\right)  r^{4}\left(  f\left(  r\right)  -1\right)  -\left(
n-3\right)  \tilde{\alpha}_{2}r^{2}\left(  f\left(  r\right)  -1\right)
^{2}+\nonumber\\
\left(  n-5\right)  \tilde{\alpha}_{3}\left(  f\left(  r\right)  -1\right)
^{3}+\left(  n-1\right)  r^{2}Q^{2}=0,
\end{gather}
in which a prime denotes derivative with respect to $r,$ $n=N-2,$
$\tilde{\alpha}_{2}=$ $\left(  n-1\right)  \left(  n-2\right)  \alpha_{2}$ and
$\tilde{\alpha}_{3}=$ $\left(  n-1\right)  \left(  n-2\right)  \left(
n-3\right)  \left(  n-4\right)  \alpha_{3}$. This equation is valid for
$N\geq4\left(  \text{i.e. }n\geq2\right)  $, but for $N=4\left(  \text{i.e.
}n=2\right)  $ we get%
\begin{equation}
r^{3}f^{\prime}\left(  r\right)  +r^{2}\left(  f\left(  r\right)  -1\right)
+Q^{2}=0,
\end{equation}
which clearly is $\alpha_{2,3}$ independent and therefore it will be the
Einstein-Yang-Mills equation admitting the well-known Reissner-Nordstrom form%
\begin{equation}
f\left(  r\right)  =1-\frac{2m}{r}+\frac{Q^{2}}{r^{2}}.
\end{equation}

For $N=5\left(  \text{i.e. }n=3\right)  $ the Eq. $\left(  18\right)  $ has
already been considered in \cite{5}. We note that in Eq. (18), $r,$
$\tilde{\alpha}_{2}$ and $\tilde{\alpha}_{3}$ can all be scaled by $Q$ (i.e.,
$r\rightarrow\left\vert Q\right\vert r,$ $\tilde{\alpha}_{2}\rightarrow
Q^{2}\tilde{\alpha}_{2}$ and $\tilde{\alpha}_{3}\rightarrow Q^{4}\tilde
{\alpha}_{3}$, as long as $Q\neq0$) so that we set in the sequel, without loss
of generality, $Q=1.$

\section{THE\ EYMGB CASE, $\alpha_{2}\neq0,$ $\alpha_{3}=0 $}

Eq. (18) with $\alpha_{3}=0$ takes the form%
\begin{gather}
\left(  r^{3}-2\tilde{\alpha}_{2}r\left(  f\left(  r\right)  -1\right)
\right)  f^{\prime}\left(  r\right)  +\\
\left(  n-1\right)  r^{2}\left(  f\left(  r\right)  -1\right)  -\left(
n-3\right)  \tilde{\alpha}_{2}\left(  f\left(  r\right)  -1\right)
^{2}+\left(  n-1\right)  =0,\nonumber
\end{gather}
which may be called as Einstein-Gauss-Bonnet-Yang-Mills (EGBYM) equation. This
equation admits a general solution in any arbitrary dimensions $N$ as follows%
\begin{equation}
f\left(  r\right)  =1+\frac{r^{2}}{2\tilde{\alpha}_{2}}\left(  1\pm
\sqrt{1+\frac{4\tilde{\alpha}_{2}m}{r^{n+1}}+\frac{4\tilde{\alpha}_{2}\left(
n-1\right)  }{\left(  n-3\right)  r^{4}}}\right)  ,\text{ \ }n>3
\end{equation}
where $m$ is the usual integration constant to be identified as mass.

\subsection{The EYMGB solution in $6-$dimensions}

In this section we shall explore some physical aspects of the solution (22) in
6-dimensions. This is interesting for the reason that, $N=5$ and $N=6$ are the
only dimensions which will not be effected by the non-zero third order
Lovelock gravity. For $N=6$ ($n=4$), the metric function $f(r)$ in Eq. (22)
takes the form%
\begin{equation}
f_{\pm}\left(  r\right)  =1+\frac{r^{2}}{2\tilde{\alpha}_{2}}\left(  1\pm
\sqrt{1+\frac{4\tilde{\alpha}_{2}m}{r^{5}}+\frac{12\tilde{\alpha}_{2}}{r^{4}}%
}\right)  ,
\end{equation}
in which $\tilde{\alpha}_{2}(=$ $6\alpha_{2})$ and $\pm$ refer to the two
different branches of the solution. Asymptotic behaviors of $f_{\pm}\left(
r\right)  $ can be shown to be as%
\begin{equation}
\underset{r\rightarrow\infty}{\lim}f_{+}\left(  r\right)  \rightarrow
1+\frac{r^{2}}{\tilde{\alpha}_{2}},\text{ \ \ and \ }\underset{r\rightarrow
\infty}{\lim}f_{-}\left(  r\right)  \rightarrow1,\nonumber
\end{equation}
which imply that, the positive branch is Asymptotically-de Sitter (A-dS) with
positive $\alpha_{2}$ and Asymptotically-Anti de Sitter (A-AdS) with negative
$\alpha_{2}.$ It is seen obviously that the negative branch is an
Asymptotically Flat (A-F) space. One can also show that
\begin{equation}
\underset{r\rightarrow0^{+}}{\lim}f_{+}\left(  r\right)  \rightarrow
+\infty,\text{\ \ \ \ and \ }\underset{r\rightarrow0^{+}}{\lim}f_{-}\left(
r\right)  \rightarrow-\infty,\nonumber
\end{equation}
which clearly, shows that, $f_{+}\left(  r\right)  $ is an A-dS solution while
$f_{-}\left(  r\right)  $ represents an A-F black hole solution. In the sequel
we shall consider $\alpha_{2}>0$ with the negative branch of the solution
(i.e., the A-F black hole solution). One can easily show that, this solution
admits a single horizon (i.e., event horizon) given by
\begin{equation}
r_{+}=\left(  \frac{m}{2}+\sqrt{\left(  \frac{m}{2}\right)  ^{2}-\left(
1-2\alpha_{2}\right)  ^{3}}\right)  ^{1/3}+\left(  \frac{m}{2}-\sqrt{\left(
\frac{m}{2}\right)  ^{2}-\left(  1-2\alpha_{2}\right)  ^{3}}\right)  ^{1/3},
\end{equation}
which is real and positive for any values of $m$ and $\tilde{\alpha}_{2}.$ In
Fig. (1), (i.e., the dashed curves) we plot the radius of event horizon
$r_{+},$ in terms of $\alpha_{2}$(i.e., Gauss-Bonnet parameter), for some
fixed values for $m.$ This figure displays the contribution of the
Gauss-Bonnet parameter to the possible radius of the event horizon of the
black hole. By looking at the Fig. (1), one may comment that for any value of
$m$, $\underset{\alpha_{2}\rightarrow\infty}{\lim}r_{+}\rightarrow0.$ We
notice further that, with $\alpha_{2}=0,$ we get the radius of event horizon
for the six dimensional EYM black hole solution which was given in the Ref.
\cite{5}. That is, for very large $\alpha_{2},$ the event horizon coincides
with the central singularity. As a particular choice, we consider $\alpha
_{2}=\frac{1}{2}$ which implies that%
\begin{equation}
r_{+}=m^{1/3}.
\end{equation}

The surface gravity, $\kappa$ defined by \cite{8}%
\begin{equation}
\kappa^{2}=-\frac{1}{4}g^{tt}g^{ij}g_{tt,i}\;g_{tt,j},
\end{equation}
takes the value%
\begin{equation}
\kappa=\left\vert \frac{1}{2}f^{\prime}\left(  r_{+}\right)  \right\vert
=\frac{3}{2}\frac{m^{1/3}}{m^{2/3}+6}.
\end{equation}

The associated Hawking temperature depending on mass $m$ and $\alpha_{2}%
=\frac{1}{2}$ becomes
\begin{equation}
T_{H}=\frac{\kappa}{2\pi}=\frac{3}{4\pi}\frac{m^{1/3}}{m^{2/3}+6},
\end{equation}
in the choice of units $c=G=\hbar=k=1.$

\subsection{The EYMGB solution in $7-$dimensions}

In this section we represent some physical aspects of the solution (22) in
7-dimensions. In 7-dimensions, both second and third order Lovelock terms
contribute but still we set $\alpha_{3}=0$ in order to identity the
contribution of $\alpha_{3}$. For $N=7$ ($n=5$), the metric function $f(r)$ in
Eq. (22) takes the form%
\begin{equation}
f_{\pm}\left(  r\right)  =1+\frac{r^{2}}{2\tilde{\alpha}_{2}}\left(  1\pm
\sqrt{1+\frac{4\tilde{\alpha}_{2}m}{r^{6}}+\frac{8\tilde{\alpha}_{2}}{r^{4}}%
}\right)  ,
\end{equation}
in which $\tilde{\alpha}_{2}(=12\alpha_{2})$ and $\pm$ \ refers to the two
individual branches of the solutions. In Fig. (2) we plot $f_{-}\left(
r\right)  $ which goes to asymptotically flat value for $r\rightarrow\infty.$
Asymptotic behaviors of $f_{\pm}\left(  r\right)  $ can by written as%
\begin{equation}
\underset{r\rightarrow\infty}{\lim}f_{+}\left(  r\right)  \rightarrow
1+\frac{r^{2}}{\tilde{\alpha}_{2}},\text{ \ \ and \ \ }\underset
{r\rightarrow\infty}{\lim}f_{-}\left(  r\right)  \rightarrow1,\nonumber
\end{equation}
which imply that, the positive branch is Asymptotically-de Sitter (A-dS) for
positive $\alpha_{2}$ and Asymptotically-Anti de Sitter (A-AdS) for negative
$\alpha_{2},$ while the negative branch leads to an Asymptotic Flat (A-F)
space. Also one can show that
\begin{equation}
\underset{r\rightarrow0^{+}}{\lim}f_{+}\left(  r\right)  \rightarrow
+\infty,\text{ \ \ and \ \ }\underset{r\rightarrow0^{+}}{\lim}f_{-}\left(
r\right)  \rightarrow-\infty,\nonumber
\end{equation}
which clearly, manifests that, $f_{+}\left(  r\right)  $ is an A-dS solution
while $f_{_{-}}\left(  r\right)  $ represents an A-F black hole solution.
Hence in the sequel we just consider $\alpha_{2}>0$ and the negative branch of
the solution (i.e., the A-F black hole solution). One can easily show that,
this solution admits only an (event horizon) which can be written as
\begin{equation}
r_{+}=\sqrt{1-6\alpha_{2}+\sqrt{\left(  1-6\alpha_{2}\right)  ^{2}+4m}},
\end{equation}
which implies that $r_{+}$ is real and positive for any values of $m$ and
$\tilde{\alpha}_{2}.$ In the Fig. (1) we plot the radius of event horizon
$r_{+},$ in terms of $\alpha_{2}$ (i.e., the solid curves), with some fixed
values of $m.$ This figure displays the contribution of the Gauss-Bonnet
parameter in place of possible radius of the event horizon of the \ EYM black
hole. We notice that, with $\alpha_{2}=0,$ we recover the radius of event
horizon for the $7-$dimensional EYM black hole solution which is given in the
Ref. \cite{7}.

As a particular choice, we consider $\alpha_{2}=1/6$ which implies that%
\begin{equation}
r_{+}=\left(  4m\right)  ^{1/4},
\end{equation}
and the surface gravity, (26) in this case has the value%
\begin{equation}
\kappa=\left\vert \frac{1}{2}f^{\prime}\left(  r_{+}\right)  \right\vert
=\frac{2m^{1/4}}{m^{1/4}+4}.
\end{equation}

With the associated Hawking temperature given by
\begin{equation}
T_{H}=\frac{\kappa}{2\pi}=\frac{1}{\pi}\frac{m^{1/4}}{m^{1/4}+4}.
\end{equation}
which is comparable with the 6-dimensional case (28).

\section{THE\ EYML CASE WITH, $\alpha_{2}=0,$ $\alpha_{3}\neq0$}

In this section we just consider the effect of $\alpha_{3}$ on the solution of
the field equation. Eq. (18) with $\alpha_{2}=0$ takes the form%
\begin{gather}
\left(  r^{5}+3\tilde{\alpha}_{3}r\left(  f\left(  r\right)  -1\right)
^{2}\right)  f^{\prime}\left(  r\right)  +\nu r^{4}\left(  f\left(  r\right)
-1\right)  +\\
\left(  \nu-4\right)  \tilde{\alpha}_{3}\left(  f\left(  r\right)  -1\right)
^{3}+\nu r^{2}=0,\nonumber
\end{gather}
in which $\nu=N-3$ ($=n-1$ therefore). The proper general solution of this
equation is given by%
\begin{equation}
f\left(  r\right)  =1+\frac{r^{4}\Omega^{\frac{1}{3}}}{6\tilde{\alpha}%
_{3}\left(  \nu-2\right)  r^{\frac{\left(  8+\nu\right)  }{3}}}-\frac{2\left(
\nu-2\right)  r^{\frac{\left(  8+\nu\right)  }{3}}}{\Omega^{\frac{1}{3}}},
\end{equation}
in which%
\[
\Omega=-108A+12\sqrt{\frac{3}{\tilde{\alpha}_{3}}\left[  27\tilde{\alpha}%
_{3}A^{2}+4\left(  \nu-2\right)  ^{2}r^{2\nu+4}\right]  },
\]
and%
\[
A=\nu r^{\nu-2}+\left(  \nu-2\right)  m.
\]

One can easily show that in the limit $\tilde{\alpha}_{3}\rightarrow0,$ we
obtain
\begin{equation}
\underset{\tilde{\alpha}_{3}\rightarrow0}{\lim}f\left(  r\right)  =1-\frac
{m}{r^{\nu}}-\frac{\nu}{\left(  \nu-2\right)  r^{2}},
\end{equation}
which is the EYM solution.

By setting $\nu=4$ in Eq. (39) for $N=7$ one obtains
\begin{equation}
f\left(  r\right)  =1+\frac{\Omega^{\frac{1}{3}}}{12\tilde{\alpha}_{3}}%
-\frac{4r^{4}}{\Omega^{\frac{1}{3}}},
\end{equation}
in which%
\begin{equation}
\Omega=-108\left(  4r^{2}+2m\right)  +12\sqrt{\frac{3}{\tilde{\alpha}_{3}%
}\left[  27\tilde{\alpha}_{3}\left(  4r^{2}+2m\right)  ^{2}+8^{2}%
r^{12}\right]  }.
\end{equation}

In Fig. (3) we plot the $f(r)$ function versus $r$ depending on different
$\tilde{\alpha}_{3}\neq0=\tilde{\alpha}_{2}$ .

\section{THE GENERAL CASE, $\alpha_{3}\neq0\neq\alpha_{2}$}

For $N\geq7$ ($n\geq5$), with $\alpha_{3}\neq0\neq\alpha_{2}$ we shall see the
role of the third order Lovelock parameters as well as the second order. This
leads us to a tedious set of differential equations which fortunately reduces
to Eq. (18) and can be integrated exactly. Indeed, the general solution of the
Eq. (18) with arbitrary values of $\alpha_{2}$ and $\alpha_{3},$ in any
arbitrary dimension $N\geq7$ can be expressed by the following expression%
\begin{align}
f\left(  r\right)   &  =1+\frac{\left(  4\Omega^{2}/\Delta\right)  ^{\frac
{1}{3}}}{6\left(  n+1\right)  \left(  n-3\right)  \tilde{\alpha}_{3}r^{n}%
}+\frac{\tilde{\alpha}_{2}r^{2}}{3\tilde{\alpha}_{3}}+\nonumber\\
&  \frac{r^{n+4}\left(  n-3\right)  \left(  n+1\right)  \left(  \tilde{\alpha
}_{2}^{2}-3\tilde{\alpha}_{3}\right)  }{6\tilde{\alpha}_{3}}\left(  \frac
{16}{\Delta}\right)  ^{\frac{1}{3}},
\end{align}
in which we use the abbreviations%
\begin{align}
\Omega &  =-r^{2n+2}\left(  n-3\right)  ^{2}\left(  1+n\right)  ^{2}\times\\
&  \left\{  -3\sqrt{\delta}\tilde{\alpha}_{3}+\left[  -\tilde{\alpha}%
_{2}\left(  2\tilde{\alpha}_{2}^{2}-9\tilde{\alpha}_{3}\right)  \left(
1+n\right)  \left(  n-3\right)  r^{4}+27\left(  n^{2}-1\right)  \tilde{\alpha
}_{3}^{2}\right]  r^{n}\right. \nonumber\\
&  +\left.  27\tilde{\alpha}_{3}^{2}mr^{3}\left(  1+n\right)  \left(
n-3\right)  \right\}  ,\nonumber
\end{align}%
\begin{align}
\delta &  =54\left(  1+n\right)  ^{2}\left\{  \frac{3}{2}\tilde{\alpha}%
_{3}^{2}\left(  n-1\right)  ^{2}r^{2n}+\right. \\
&  \left[  \left(  n-1\right)  \left(  n-3\right)  \tilde{\alpha}_{2}\left(
\tilde{\alpha}_{3}-\frac{2}{9}\tilde{\alpha}_{2}^{2}\right)  r^{2n+4}%
+3\tilde{\alpha}_{3}^{2}m\left(  n-1\right)  \left(  n-3\right)
r^{n+3}\right] \nonumber\\
&  +\frac{2}{9}\left(  \tilde{\alpha}_{3}-\frac{1}{4}\tilde{\alpha}_{2}%
^{2}\right)  \left(  n-3\right)  ^{2}r^{8+2n}+\left(  \tilde{\alpha}_{3}%
-\frac{2}{9}\tilde{\alpha}_{2}^{2}\right)  \left(  n-3\right)  ^{2}%
\tilde{\alpha}_{2}mr^{7+n}\nonumber\\
&  +\left.  \frac{3}{2}\tilde{\alpha}_{3}^{2}mr^{6}\left(  n-3\right)
^{2}\right\}  ,\nonumber
\end{align}%
\begin{align}
\Delta &  =2\left(  n-3\right)  ^{2}\left\{  \frac{3}{2}\sqrt{\delta}%
\tilde{\alpha}_{3}+\left(  1+n\right)  \times\right. \\
&  \left[  \left(  -\frac{27}{2}\left(  n-1\right)  \tilde{\alpha}_{3}%
^{2}-\frac{9}{2}\tilde{\alpha}_{2}\tilde{\alpha}_{3}r^{4}\left(  n-3\right)
+\tilde{\alpha}_{2}^{3}r^{4}\left(  n-3\right)  \right)  r^{n}\right.
\nonumber\\
&  \left.  \left.  -\frac{27}{2}\tilde{\alpha}_{2}^{3}mr^{3}\left(
n-3\right)  \right]  \right\}  r^{2n+2}\left(  1+n\right)  ^{2}.\nonumber
\end{align}

To proceed further with these expressions doesn't seem feasible from technical
points, therefore in the sequel we shall adopt a special simplifying relation
between $\alpha_{2}$ and $\alpha_{3}$.

\subsection{The EYML solution in $7-$dimensions with $\tilde{\alpha}%
_{3}=\tilde{\alpha}_{2}^{2}/3$}

In $7-$dimensional space time we can see the roles of both second and third
order Lovelock parameters simultaneously. One of the simplifying, yet
interesting case in the solution (43) can be obtained if one sets
$\tilde{\alpha}_{3}=\frac{\tilde{\alpha}_{2}^{2}}{3}.$ The metric function
$f(r)$ reads rather simple%
\begin{equation}
f\left(  r\right)  =1+\frac{r^{2}}{\tilde{\alpha}_{2}}\left\{  1-\left[
1+\frac{3\tilde{\alpha}_{2}m}{r^{n+1}}+\frac{3\left(  n-1\right)
\tilde{\alpha}_{2}}{\left(  n-3\right)  r^{4}}\right]  ^{\frac{1}{3}}\right\}
,
\end{equation}
which yields an asymptotically flat metric. For $r\rightarrow\infty$ in order
to have black hole solutions one should investigate the existence of roots of
the metric function (i.e., $f(r)=0$). To this end, one finds the solutions of
the following equation%
\begin{equation}
r^{n+1}\left[  r^{4}+\left(  \tilde{\alpha}_{2}-\frac{n-1}{n-3}\right)
r^{2}+\frac{1}{3}\tilde{\alpha}_{2}^{2}\right]  -mr^{6}=0,
\end{equation}
\bigskip which generally seems a difficult task. But in seven dimensions we
get the following roots%
\begin{equation}
r_{\pm}=\left[  \left(  1-\frac{\tilde{\alpha}_{2}}{2}\right)  \pm
\sqrt{\left(  1-\frac{\tilde{\alpha}_{2}}{2}\right)  ^{2}+\left(
m-\frac{\tilde{\alpha}_{2}^{2}}{3}\right)  }\right]  ^{\frac{1}{2}},
\end{equation}
in which for $1>\sqrt{\frac{\tilde{\alpha}_{2}}{2}}$ and $\left[  \frac
{\tilde{\alpha}_{2}^{2}}{3}-\left(  1-\frac{\tilde{\alpha}_{2}}{2}\right)
^{2}\right]  <m<\frac{\tilde{\alpha}_{2}^{2}}{3},$ so that we have both inner
and outer radii of black hole. In the upper and lower limits of $m,$ we will
get just the radius of the event horizon of the black hole solutions, i.e.,
when $m=\frac{\tilde{\alpha}_{2}^{2}}{3},$ then $r_{+}=\sqrt{2-\tilde{\alpha
}_{2}}$ while $r_{-}=0.$ On the other hand when $m=\frac{\tilde{\alpha}%
_{2}^{2}}{3}-\left(  1-\frac{\tilde{\alpha}_{2}}{2}\right)  ^{2},$
$r_{+}=\sqrt{1-\frac{\tilde{\alpha}_{2}}{2}}$ while $r_{-}$ does not exist.
For the choice $\tilde{\alpha}_{2}=2,$ we can define a critical mass,
$m_{crit}=4/3,$ so that for $m>m_{crit}$ we can have a black hole solution.

The surface gravity in $N=7$ has the form
\begin{equation}
\kappa=\left\vert \frac{1}{2}f^{\prime}\left(  r_{+}\right)  \right\vert
=\left\vert \frac{r_{+}}{\tilde{\alpha}_{2}}\left(  1-\frac{1}{\Delta^{\left(
2/3\right)  }}\right)  -\frac{2}{r_{+}^{3}\Delta^{\left(  2/3\right)  }%
}\right\vert \text{ ,\ }%
\end{equation}
in which
\[
\Delta=1+\frac{3\tilde{\alpha}_{2}m}{r_{+}^{6}}+\frac{6\tilde{\alpha}_{2}%
}{r_{+}^{4}},
\]
where $r_{+}$ is the radius of event horizon of the possible black hole. For
instance when $\tilde{\alpha}_{2}=2,$ (i.e., $r_{+}=\left(  m-\frac{4}%
{3}\right)  ^{1/4}$) one gets%
\begin{equation}
\kappa=\left\vert 9\sqrt{3}\frac{\left(  3m-4\right)  ^{2}\left[  \left(
3m-4\right)  \Xi^{2/3}-\left(  3m+4\right)  \right]  }{2\left(  9m-12\right)
^{11/4}\Xi^{2/3}}\right\vert ,
\end{equation}
in which%
\[
\Xi=1+\frac{162m}{\left(  9m-12\right)  ^{3/2}}+\frac{36}{\left(  3m-4\right)
}.
\]

The associated Hawking temperature $T_{H}=\frac{\kappa}{2\pi}$ can be found by
using the above result, on which one may expect that, $T_{H}$ is strongly
$\tilde{\alpha}_{2}$ (and $\tilde{\alpha}_{3}$) dependent.

\subsection{Asymptotically dS (AdS) property}

The general solution (39) as $r\rightarrow\infty$ reads%
\begin{equation}
f_{\infty}\left(  r\right)  \tilde{=}1+\left[  \left(  \frac{\tilde{\alpha
}_{2}^{2}}{3\tilde{\alpha}_{3}}-1\right)  \left(  \frac{2}{\Sigma}\right)
^{1/3}+\frac{1}{3\tilde{\alpha}_{3}}\left(  \frac{\Sigma}{2}\right)
^{1/3}+\frac{\tilde{\alpha}_{2}}{3\tilde{\alpha}_{3}}\right]  r^{2},
\end{equation}
where
\[
\Sigma=\left[  3\tilde{\alpha}_{3}\sqrt{3\left(  4\tilde{\alpha}_{3}%
-\tilde{\alpha}_{2}^{2}\right)  }+2\tilde{\alpha}_{2}^{3}-9\tilde{\alpha}%
_{2}\tilde{\alpha}_{3}\right]  .
\]

It is observed that the metric function (39), can be rewritten as a $N$
\ ($n=N-2$) dimensional dS space in which $f_{\infty}\left(  r\right)
\tilde{=}1-\frac{2\tilde{\Lambda}}{\left(  n-3\right)  \left(  n-4\right)
}r^{2}$ one defines a cosmological constant without cosmological
constant$-\tilde{\Lambda}$ as an effective cosmological constant given by
\begin{equation}
\tilde{\Lambda}=-\frac{\left(  n-3\right)  \left(  n-4\right)  }{2}\left[
\left(  \frac{\tilde{\alpha}_{2}^{2}}{3\tilde{\alpha}_{3}}-1\right)  \left(
\frac{2}{\Sigma}\right)  ^{1/3}+\frac{1}{3\tilde{\alpha}_{3}}\left(
\frac{\Sigma}{2}\right)  ^{1/3}+\frac{\tilde{\alpha}_{2}}{3\tilde{\alpha}_{3}%
}\right]  .
\end{equation}

It is seen that in this effective cosmological constant both $\alpha_{2},$ and
$\alpha_{3}$ play role. One can easily show that, depending on $\tilde{\alpha
}_{2}$ and $\tilde{\alpha}_{3},$ $\tilde{\Lambda}$ can take zero, positive or
negative values, and consequently the general solution becomes asymptotically
flat, dS or AdS, respectively. For instance, from Eq. (43), one can show that
the choice $\tilde{\alpha}_{3}=\tilde{\alpha}_{2}^{2}/3$ is asymptotically flat.

These results verify that, the Lovelock parameters $\tilde{\alpha}_{2}$ and
$\tilde{\alpha}_{3}$ significantly modify the properties of EYM black holes as
well as their asymptotic behaviors.

\section{CONCLUSION}

We introduced YM fields through the Wu-Yang Ansatz into the third order
Lovelock gravity with spherical symmetry. The Ansatz and symmetry aided in
overcoming the technical difficulty and obtaining exact solutions in higher
dimensions. In this sense our work is partly an extension of our previous work
which included only the GB parameter and for $N=5$ \cite{5}. Our solutions
include black hole possessing parameters of mass, magnetic charge (which is
scaled to Q=1), $\alpha_{2}$ (for $N\geq5)$ and $\alpha_{3}$ (for $N\geq7)$.
Our analysis indicates that higher order Lovelock parameters have less
significant contributions compared with the lower order terms superposed to
the EH Lagrangian. At least this is the picture that we expect in case that
$\alpha_{k}$ $(k\geq4)$ terms are taken into account. Depending on the
choice/relative magnitudes of the parameters, formation of the black holes
with inner/outer horizons is conditional. Asymptotic solutions give rise to dS
(AdS) or flat space times which are topologically trivial but from the field
theoretic point of view they are rather important.

\section{Figure Captions:}

Fig. (1): The horizon radius $r_{h}$ versus $\tilde{\alpha}_{2}$ for
$\tilde{\alpha}_{3}=0$ , and different values for mass (each mass is written
on the correspondence curve) in N=6 and N=7.

Fig. (2): Plot of the metric function $f\left(  r\right)  $ versus $r$ for
N=7, m=1, $\tilde{\alpha}_{3}=0,$ $\tilde{\alpha}_{2}=5,$ $1,$ $0.5,$ $0.1,$
$0.01$ and $0.001.$

Fig. (3): Plot of the metric function $f(r)$ versus $r$ for N=7, m=1,
$\tilde{\alpha}_{2}=0,$ $\tilde{\alpha}_{3}=5,$ $1,$ $0.5,$ $0.1,$ $0.01$ and
$0.001.$

\end{document}